\documentclass[review]{elsarticle}
\usepackage{rotating}

\journal{Journal of \LaTeX\ Templates}

\begin{document}

\begin{frontmatter}

\title{Charge Transport Behavior of 1D Gold Chiral Nanojunctions}

\author{Talem Rebeda Roy$^{a,b}$, Arijit Sen$^{a,b*}$}
\address{$^{a}$SRM Research Institute, SRM Institute of Science and Technology, Chennai 603203, India}
\address{$^{b}$Department of Physics and Nanotechnology, SRM Institute of Science and Technology, Chennai 603203, India}


\cortext[mycorrespondingauthor]{Corresponding author}
\ead{arijit.s@res.srmuniv.ac.in}


\begin{abstract}
Understanding the process of electron tunneling in chirality-induced single-molecule junctions is imperative for the development of nanoscale switching and artificial nanomotors. Based on the combined non-equilibrium Green’s functions formalism and the ground-state density functional theory, we present here the charge transport behavior of $chiral$ gold (7,3) nanowires (NWs) in comparison with various other $chiral$ and $achiral$ 1D gold nanostructures as the principal leads to form stable single-molecule junctions. For $\sigma$-saturated alkane chains, we find that the contact potential barriers vary widely with the $achiral$ leads but not with the $chiral$ ones, although a close resemblance exists in the tunneling constants. Lower energy gaps for single-molecule junctions with Au(7,3)NWs ensure better electronic conductance even after allowing for the low thermal loss, due mainly to the close-packed arrangements of atoms with minimum wire tension. Our $first-principles$ quantum transport analysis further suggests that $chiral$ Au(7,3)NWs render higher electronic conductance than $chiral$ gold (5,3) nanotubes (NTs), once bridged by either $\sigma$-saturated or $\pi$-conjugated molecular moieties. It, however, turns out that asymmetricity in the characteristics of channel formation at the lead-molecule contact remains often associated with $chiral$ Au(7,3)NWs only.
\end{abstract}


\end{frontmatter}


\section{Introduction}
Recent years have seen a sizeable amount of study on one-dimensional (1D) nanostructures in the form of nanotubes (NTs) and nanowires (NWs), primarily because of their potential applications in nanodevice technology \cite{1shyu,2geetha,3gao,4oshima,5tosatti,6senger,7lambert,8valle,9oleg,10man}. Among these, $chiral$ nanostructures continue to garner interest due especially to their characteristic helicity that often displays the ability to control the physical properties in somewhat more flexible way. From conformational point of view, $achiral$ nanostructures possess mirror planes while $chiral$ nanostructures display no mirror symmetry but have glide planes. Magnetoelectronic properties are often found to be quite sensitive to chirality \cite{1shyu}, apart from the band-related differences like band symmetry and band spacing between $chiral$ and $achiral$ carbon nanotubes (CNTs). Further, the impurity concentration and curvature effect turn out to be much stronger in $chiral$ CNTs than in $achiral$ ones\cite{2geetha}.

Although a major focus has remained for quite some time on developing CNT based field effect transistors (CNT-FETs) in a robust way, metallic nanotubes and nanowires have also received much attention\cite{4oshima,5tosatti,6senger,7lambert,8valle,9oleg} in the process. The underlying lattice for a gold nanotube is a 2D triangular network with one atom per unit cell, in contrast to a CNT that has a hexagonal framework. Another important difference is that gold nanotubes, unlike CNTs, always display metalicity due to its unique electronic nature. Using a tight-binding spiral model, Yevtushenko \textit{et al} \cite{9oleg} have demonstrated that electronic transport may get affected considerably by the interplay between chirality and nonlinearity in a $chiral$ CNT. However, the works by Manrique \textit{et al} \cite{10man} reveal that like in CNTs, $chiral$ currents are also oscillatory functions of energy in Au(5,3)NTs, but the magnitude of $chiral$ currents are much larger in the latter. The first experimental observation of $chiral$ Au(5,3)NT based nanobridges was reported by Oshima \textit{et al}\cite{4oshima}. Later, Senger \textit{et al}\cite{6senger} predicted that such conformation would possess a “magic” size\cite{5tosatti,6senger} in respect of its minimum wire tension. We have recently shown that $chiral$ Au(5,3)NTs can form stable single-molecule junctions having similar conductance as $achiral$ Au(100)NWs\cite{11sen}. In this work, we intend to address how the metallic chirality, as shown in Fig. 1, may affect the electronic transport in respect of $achiral$ counterparts while forming a single-molecule junction. 

\section{Computational methods}
Several two probe junctions formed by leads of $chiral$ as well as $achiral$ symmetry were constructed being abridged by either a $\sigma$-saturated alkane chain of varying length or a $\pi$-conjugated benzene ring. Leads were also chosen as either gold nanotubes (AuNTs) or gold nanowires (AuNWs) having $chiral$ or $achiral$ character. All these leads were pre-optimized using the densiy functional theory (DFT) before the formation of respective two probe junctions. The lattice parameters along the x and y directions were maintained at a minimum of 24 $\AA$ to allow for sufficient isolation during the 1D relaxation. The optimized lattice constant along the z direction in the unit cell turns out to be 4.08, 7.06, 20.73 and 28.55 $\AA$ respectively for Au(100)NW, Au(111)NW, Au(5,3)NT and Au(7,3)NW based leads. The number of atoms in the respective unit cell varies from as minimum as 9 for $achiral$ Au(100)NW to as maximum as 85 for $chiral$ AU(7,3)NW. The molecular moieties in all these heterojunctions were optimized within a force tolerance limit of $0.05eV/\AA$, with a basis of linear combination of atomic orbitals (LCAO), as implemented in the SIESTA package\cite{SIESTA2002}.

For the transport calculations, we employed the non-equilibrium Green’s functions (NEGF)\cite{12taylor,13bran,Liu2016} formalism by utilizing non-orthogonal localized $double-\zeta$ basis sets, while core electrons were treated by norm-conserving pseudopotentials. In our NEGF-DFT calculations\cite{chen2012,chen2012210,zhang2017,hu2017}, the density mesh cutoff was set to 200 Ry and the $k$-point sampling grid was chosen as 1$\times$1$\times$400. The exchange and correlation were handled within the generalized gradient approximation (GGA) based on the revised Perdew, Burke, and Ernzerhof (rPBE) functional\cite{rPBE1998}. The transmission probability, $T(E,V_{b})$, at a given energy (E) and bias ($V_{b}$), can be obtained from the generalized Landauer theory\cite{14landauer} as
\begin{equation}
T(E,V_{b})=Tr[\Gamma_{L}(E)G^{r}(E)\Gamma_{R}(E)G^{a}(E)]
\end{equation}
Where $G^{r(a)}$  denote the  retarded (advanced) Green’s functions of the scattering region and $\Gamma_{L(R)}=\Im\sum_{L(R)}^{r}(E)-\sum_{L(R)}^{r^{\dagger}}(E) $ signify the line widths due to the coupling of the molecule, $\sum_{L(R)}^{r}(E)$, with the left (L) and right (R) leads, as determined from the unperturbed lead Green’s functions. The conductances were obtained from the transmission at the Fermi level such that $G=G_{0}\sum_{i}T_{i} $, where $G_{0}$ is the conductance quantum\cite{14landauer} while $T_{i}$ refers to the eigenvalue of the $i$-th eigenchannel.

\section{Results and discussion}
From the bandstructure analyses, we know that Au(5,3)NT and Au(7,3)NW possess respectively five and six available channels\cite{6senger} for conduction leading to the electronic  conductance values of $5G_{0}$ and $6G_{0}$. The fact that the number of helical strands in Au(7,3)NW exceeds that of Au(5,3)NT leads to better conductance in the former [see Table 1]. However, when these 1D nanostructures, acting as leads, form single-molecule junctions with the same active element, appreciable diversity crops up in the behavior of electron transmission. Even as the length of the alkanedithiol $[ADT, HS-(CH_{2})_{N}-SH]$ molecular chain increases, the resonant transmission peaks undergo noticeable reduction in amplitude in case of Au(5,3)NTs, working as leads (see Fig. 2). Further, Au(100)NW provides the high conductance (HC) while Au(111)NW, the low conductance (LC) values even for the increasing molecular length, which is in tune with our previous work \cite{15sen}. In contrast, 1D  gold $chiral$ nanostructures, irrespective of being NTs or NWs, exhibit only HC in alkanedithiol SMJs. As Table 1 suggests, our calculated HC and LC values auger well with the available experimental data.
\begin{sidewaystable}
	\small
	\caption{\ A comparative study of the conductance for various $achiral$ and $chiral$ single-molecule junctions as the length of the $\sigma$-saturated alkanedithiol molecular chain is increased in terms of the number (N) of the methylene groups. The available experimental data are also listed, associated with Au clusters being leads.}
	\label{tbl:example}
	\begin{tabular*}{\textwidth}{@{\extracolsep{\fill}}|c|cc|cc|}
		\cline{1-5}
		&\multicolumn{4}{c|}{Conductance ($G_{0}$)  Cal. (Exp.)}\\
		\cline{2-5}
		N&\multicolumn{2}{c|}{Achiral}&\multicolumn{2}{c|}{chiral}\\
		\cline{2-5}
		&Au(100)NW&Au(111)NW&Au(5,3)NT&Au(7,3)NW\\
		\cline{1-5}
		4&	1.0$\times10^{-2}$	&1.3$\times10^{-3}$&	1.0$\times10^{-2}$	&1.5$\times10^{-2}$\\
		6&	1.0$\times10^{-3}$ (1.0$\times10^{-3}$)\cite{16fu}&	2.0$\times10^{-4}$ (1.8$\times10^{-4}$)\cite{16fu}&	1.6$\times10^{-3}$ (1.4$\times10^{-3}$)\cite{16fu}&	3.3$\times10^{-3}$\\
		8&	1.7$\times10^{-4}$ (2.2$\times10^{-4}$)\cite{16fu}&	2.7$\times10^{-5}$&	1.9$\times10^{-4}$&	3.6$\times10^{-4}$ ( 3.0$\times10^{-4}$)\cite{17li}\\
		10&	2.5$\times10^{-5}$ (2.0$\times10^{-5}$)\cite{17li}&	4.0$\times10^{-6}$ (4.0$\times10^{-6}$, 5.7$\times10^{-6}$)\cite{16fu,18jang}&	2.8$\times10^{-5}$ (2.6$\times10^{-5}$)\cite{16fu}&	4.5$\times10^{-5}$\\
		\cline{1-5}
	
	\end{tabular*}
\end{sidewaystable}

From Fig. 2(a-d), we find that the transmission peak for $achiral$ Au(100)NW based SMJs at -0.22 eV remains strong even as the length of the alkanedithiol molecular wire increases. It arises from the band edge states of Au(100)NW leading to Fano resonance through quantum interference, mediated by the through-bond as well as the through-space tunneling events\cite{15sen}.  However, the other peak near the Fermi level at about -0.1 eV appears in the form of Breit-Wigner resonance due to effective coupling between the lead and the respective alkane chain. Such coupling indeed varies as the lead topology and also, the molecular conformation tend to differ. To understand this, we calculate the renormalized molecular levels (RMLs) from the diagonalization of the self-consistent Hamiltonian projected on the molecular moiety\cite{12taylor}. A similar trend is observed for $achiral$ Au(111)NW based SMJs as well at, however, different electronic energies. 
A transmission peak would arise whenever there is an available state of similar orbital character in the leads, which can interact with the RMLs in order to yield the resonance. As exhibited in Fig. 2(e-h), the two resonant transmission peaks near the Fermi level for $chiral$ Au(5,3)NT based SMJs at respectively -0.1 and -0.2 eV disappear as the number (N) of the methylene groups  increases beyond N=6. This is because the lead-molecule coupling gets weakened abruptly with the increase in the molecular length. Contrary to it, the resonant transmission peak at about -0.06 eV, associated with $chiral$ Au(7,3)NW based SMJs, remains prominent even up to N=10, though gets reduced as usual in amplitude. However, the primary lead-molecule interaction in all these heterojunctions has $\pi$ character, stemming from the $p_{x}$ orbital of S linkers and the $d_{xz}$ orbital of Au adatoms.
Contour plots in the inset of Fig. 3 for the local density of states (LDOS) at the Fermi level imply quantum tunneling as the principal mechanism of charge transport in the $chiral$ Au(7,3)NW system as well. The through-space tunneling strength\cite{15sen,19papa} begins to weaken as the tunneling length expands because of the exponential behavior in the electronic conductance (G) with the increasing molecular length (L), as described by $G⁄G_{0}=A_{G} exp(-\beta_{L} L)$, where $\beta_{L}$refers to the tunneling decay parameter and $ln A_{G}$, the intercept. From the plot of natural logarithm of electronic conductance versus molecular length for a set of alkanedithiol SMJs associated with 1D gold $chiral$ and $achiral$ nanostructures, as shown in  Fig. 3, it is apparent that $chiral$ nanoleads yield the high conductance (HC), while $achiral$ ones out of Au(111)NW are only responsible for the low conductance (LC). As we see in Table 2, the tunneling decay parameter does not depend much on chirality though the intercepts as well as the contact resistance vary a lot. To further analyze the contact properties, we calculated the contact decay constant ($\beta_{C}$) and also, the contact potential barrier height ($\phi_{C}$), according to Ref. 27. It turns out that the available experimental values corroborate often to the $chiral$ systems only, as far as the contact properties are concerned\cite{20zhou}.
\begin{table*}[htb!]
	\small
	\caption{\  A comparative study of the decay constant ($\beta_{L}$), the intercept ($ln A_{G}$), the contact conductance ($G_{C}$), the contact decay constant ($\beta_{C}$) and the contact potential barrier height ($\phi_{C}$) for various $achiral$ as well as $chiral$ single-molecule junctions, as obtained from our first-principles analysis. The available experimental data,associated with Au clusters being the leads, are also listed.}
	
	\label{tbl:example}
	\begin{tabular*}{\textwidth}{@{\extracolsep{\fill}}|c|cc|cc|c|}
		\cline{1-6}
		Contact properties&	\multicolumn{2}{c|}{Achiral}	&\multicolumn{2}{c|}{Chiral}&	Au cluster ($Exp.$)\\
		\cline{2-5}
	&Au(100)NW\cite{15sen}&	Au(111)NW\cite{15sen}&	Au(5,3)NT&	Au(7,3)NW&\\
	\cline{1-6}	
	$\beta_{L}    (\AA^{-1})$&	1.1&	0.9&	1.0&	0.9 & 0.84$\pm$0.04\cite{21zhou}\\
	$In A_{G}$&	-0.10&	-2.78&	-0.47&	-0.46&	-0.43\cite{21zhou}\\
	$G_{C}       (G_{0})$&	1.11&	0.062&	0.62&	0.63	&0.65\cite{21zhou}\\
	$\beta_{C}     (\AA^{-1})$&	0.023&	0.63&	0.1&	0.1	&0.115$\pm$0.035\cite{20zhou}\\
	$\phi_{C}       (eV)$&	0.001&	0.39&	0.01&	0.01&	0.01\cite{20zhou}\\
	\cline{1-6}
		
	\end{tabular*}
\end{table*}

We further examine the behavior of charge transport in single-molecule junctions with respect to the benzenedithiol (BDT) molecule. As Fig. 4(a-d) shows, the electronic conductance increases manifold, in sharp contrast to alkanedithiol SMJs, irrespective of the lead chirality. Such enhancement occurs because of the channel enhancement with the availability of $\pi$-conjugated molecular orbitals. It is further evident from the respective transmission eigenstates having the isovalue of 0.4, associated with the principal resonant peaks near the Fermi level. The large width in the transmission peaks of $achiral$ Au(100)NW based SMJs signify the onset of strong lead-molecule coupling while such coupling appears to be weaker with $chiral$ systems. From Fig. 4(c), the single resonant peak at the lowest unoccupied molecular orbital (LUMO) of the $chiral$ Au(5,3)NT based SMJs implies the LUMO-mediated electron transmission. On the other hand, at the highest occupied molecular orbital (HOMO) of the $chiral$ Au(7,3)NW based SMJs signifies the HOMO-mediated transmission, as indicated in Fig. 4(d). However, in all these heterojunctions, $\sigma$ channel appears to be better coupled through $d_{z^{2}}$ orbital of the Au adatoms. 

\section{Conclusions}
We have demonstrated from $first-principles$ the charge transport behavior in various stable single-molecule junctions, made up of $chiral$ as well as $achiral$ leads at the nanoscale. Apparently, there lies a considerable difference in the transmission pattern when a single molecule is trapped between leads of  diverse topology in forming a break junction. While single-molecule junctions with $chiral$ leads like Au(5,3)NTs and Au(7,3)NWs give a close resemblance to those with $achiral$ leads like Au(100)NWs in the length dependence of electronic conductance, the contact potentials differ appreciably from $chiral$ to $achiral$ heterojunctions and sometimes, even within $achiral$ systems as well. For $\sigma$-saturated alkane chains, Au(7,3)NWs give rise to higher electronic conductance with asymmetric lead-molecule contact than Au(5,3)NTs that display rather symmetricity in the $\pi$-channel formation. On the other hand, single-molecule junctions having $\pi$-conjugated molecular moieties bridging Au(5,3)NTs render lower electronic conductance out of the LUMO-mediated transmission than those linking Au(7,3)NWs as leads following the HOMO-mediated one, though with strong $\sigma$-channel coupling. Understanding the charge transport mechanism of stable single-molecule junctions having $chirality$ features at the nanoscale leads can potentially help develop molecular switching and artificial nanomotors, much of which exist still at the laboratory level only.

\section{Acknowledgement}
This work was supported by DST NanoMission, Govt. of India, $via$ Project No. SR/NM/NS1062/2012. We are thankful to the National PARAM Supercomputing Facility (NPSF), Centre for Development of Advanced Computing (C-DAC), along with SRM-HPCC, for facilitating the high-performance computing.

\section*{References}

\bibliography{mybibfile}

\begin{thebibliography}{10}
\expandafter\ifx\csname url\endcsname\relax
  \def\url#1{\texttt{#1}}\fi
\expandafter\ifx\csname urlprefix\endcsname\relax\def\urlprefix{URL }\fi
\expandafter\ifx\csname href\endcsname\relax
  \def\href#1#2{#2} \def\path#1{#1}\fi

\bibitem{1shyu}
F.~L. Shyu, C.~C. Tsai, C.~H. Lee, M.~F. Lin,
  \href{http://stacks.iop.org/0953-8984/18/i=35/a=016}{Magnetoelectronic
  properties of chiral carbon nanotubes and tori}, Journal of Physics:
  Condensed Matter 18~(35) (2006) 8313.
\newline\urlprefix\url{http://stacks.iop.org/0953-8984/18/i=35/a=016}

\bibitem{2geetha}
R.~Geetha, V.~Gayathri,
  \href{http://stacks.iop.org/1402-4896/80/i=2/a=025701}{Strong effect of
  chirality on doped single walled carbon nanotubes}, Physica Scripta 80~(2)
  (2009) 025701.
\newline\urlprefix\url{http://stacks.iop.org/1402-4896/80/i=2/a=025701}

\bibitem{3gao}
P.~Gao, H.~Li, Q.~Zhang, N.~Peng, D.~He,
  \href{http://stacks.iop.org/0957-4484/21/i=9/a=095202}{Carbon nanotube
  field-effect transistors functionalized with self-assembly gold
  nanocrystals}, Nanotechnology 21~(9) (2010) 095202.
\newline\urlprefix\url{http://stacks.iop.org/0957-4484/21/i=9/a=095202}

\bibitem{4oshima}
Y.~Oshima, A.~Onga, K.~Takayanagi,
  \href{https://link.aps.org/doi/10.1103/PhysRevLett.91.205503}{Helical gold
  nanotube synthesized at 150 k}, Phys. Rev. Lett. 91 (2003) 205503.
\newblock \href {http://dx.doi.org/10.1103/PhysRevLett.91.205503}
  {\path{doi:10.1103/PhysRevLett.91.205503}}.
\newline\urlprefix\url{https://link.aps.org/doi/10.1103/PhysRevLett.91.205503}

\bibitem{5tosatti}
E.~Tosatti, S.~Prestipino, S.~Kostlmeier, A.~D. Corso, F.~D. Di~Tolla,
  \href{http://science.sciencemag.org/content/291/5502/288}{String tension and
  stability of magic tip-suspended nanowires}, Science 291~(5502) (2001)
  288--290.
\newblock \href
  {http://arxiv.org/abs/http://science.sciencemag.org/content/291/5502/288.full.pdf}
  {\path{arXiv:http://science.sciencemag.org/content/291/5502/288.full.pdf}},
  \href {http://dx.doi.org/10.1126/science.291.5502.288}
  {\path{doi:10.1126/science.291.5502.288}}.
\newline\urlprefix\url{http://science.sciencemag.org/content/291/5502/288}

\bibitem{6senger}
R.~T. Senger, S.~Dag, S.~Ciraci,
  \href{https://link.aps.org/doi/10.1103/PhysRevLett.93.196807}{Chiral
  single-wall gold nanotubes}, Phys. Rev. Lett. 93 (2004) 196807.
\newblock \href {http://dx.doi.org/10.1103/PhysRevLett.93.196807}
  {\path{doi:10.1103/PhysRevLett.93.196807}}.
\newline\urlprefix\url{https://link.aps.org/doi/10.1103/PhysRevLett.93.196807}

\bibitem{7lambert}
C.~J. Lambert, S.~W.~D. Bailey, J.~Cserti,
  \href{https://link.aps.org/doi/10.1103/PhysRevB.78.233405}{Oscillating chiral
  currents in nanotubes: A route to nanoscale magnetic test tubes}, Phys. Rev.
  B 78 (2008) 233405.
\newblock \href {http://dx.doi.org/10.1103/PhysRevB.78.233405}
  {\path{doi:10.1103/PhysRevB.78.233405}}.
\newline\urlprefix\url{https://link.aps.org/doi/10.1103/PhysRevB.78.233405}

\bibitem{8valle}
M.~del Valle, C.~Tejedor, G.~Cuniberti,
  \href{https://link.aps.org/doi/10.1103/PhysRevB.74.045408}{Scaling of the
  conductance in gold nanotubes}, Phys. Rev. B 74 (2006) 045408.
\newblock \href {http://dx.doi.org/10.1103/PhysRevB.74.045408}
  {\path{doi:10.1103/PhysRevB.74.045408}}.
\newline\urlprefix\url{https://link.aps.org/doi/10.1103/PhysRevB.74.045408}

\bibitem{9oleg}
O.~M. Yevtushenko, G.~Y. Slepyan, S.~A. Maksimenko, A.~Lakhtakia, D.~A.
  Romanov,
  \href{https://link.aps.org/doi/10.1103/PhysRevLett.79.1102}{Nonlinear
  electron transport effects in a chiral carbon nanotube}, Phys. Rev. Lett. 79
  (1997) 1102--1105.
\newblock \href {http://dx.doi.org/10.1103/PhysRevLett.79.1102}
  {\path{doi:10.1103/PhysRevLett.79.1102}}.
\newline\urlprefix\url{https://link.aps.org/doi/10.1103/PhysRevLett.79.1102}

\bibitem{10man}
D.~Z. Manrique, J.~Cserti, C.~J. Lambert,
  \href{https://link.aps.org/doi/10.1103/PhysRevB.81.073103}{Chiral currents in
  gold nanotubes}, Phys. Rev. B 81 (2010) 073103.
\newblock \href {http://dx.doi.org/10.1103/PhysRevB.81.073103}
  {\path{doi:10.1103/PhysRevB.81.073103}}.
\newline\urlprefix\url{https://link.aps.org/doi/10.1103/PhysRevB.81.073103}

\bibitem{11sen}
A.~Sen, C.-J. Lin, C.-C. Kaun,
  \href{http://dx.doi.org/10.1021/jp402531p}{Single-molecule conductance
  through chiral gold nanotubes}, The Journal of Physical Chemistry C 117~(26)
  (2013) 13676--13680.
\newblock \href {http://arxiv.org/abs/http://dx.doi.org/10.1021/jp402531p}
  {\path{arXiv:http://dx.doi.org/10.1021/jp402531p}}, \href
  {http://dx.doi.org/10.1021/jp402531p} {\path{doi:10.1021/jp402531p}}.
\newline\urlprefix\url{http://dx.doi.org/10.1021/jp402531p}

\bibitem{SIESTA2002}
J.~M. Soler, E.~Artacho, J.~D. Gale, A.~García, J.~Junquera, P.~Ordejón,
  D.~Sánchez-Portal, \href{http://stacks.iop.org/0953-8984/14/i=11/a=302}{The
  siesta method for ab initio order- n materials simulation}, Journal of
  Physics: Condensed Matter 14~(11) (2002) 2745.
\newline\urlprefix\url{http://stacks.iop.org/0953-8984/14/i=11/a=302}

\bibitem{12taylor}
J.~Taylor, H.~Guo, J.~Wang,
  \href{https://link.aps.org/doi/10.1103/PhysRevB.63.245407}{Ab initio}, Phys.
  Rev. B 63 (2001) 245407.
\newblock \href {http://dx.doi.org/10.1103/PhysRevB.63.245407}
  {\path{doi:10.1103/PhysRevB.63.245407}}.
\newline\urlprefix\url{https://link.aps.org/doi/10.1103/PhysRevB.63.245407}

\bibitem{13bran}
M.~Brandbyge, J.-L. Mozos, P.~Ordej\'on, J.~Taylor, K.~Stokbro,
  \href{https://link.aps.org/doi/10.1103/PhysRevB.65.165401}{Density-functional
  method for nonequilibrium electron transport}, Phys. Rev. B 65 (2002) 165401.
\newblock \href {http://dx.doi.org/10.1103/PhysRevB.65.165401}
  {\path{doi:10.1103/PhysRevB.65.165401}}.
\newline\urlprefix\url{https://link.aps.org/doi/10.1103/PhysRevB.65.165401}

\bibitem{Liu2016}
N.~Liu, L.~Zhang, X.~Chen, X.~Kong, X.~Zheng, H.~Guo,
  \href{http://dx.doi.org/10.1039/C6NR05087E}{Negative differential resistance
  in gesi core-shell transport junctions: the role of local sp2 hybridization},
  Nanoscale 8 (2016) 16026--16033.
\newblock \href {http://dx.doi.org/10.1039/C6NR05087E}
  {\path{doi:10.1039/C6NR05087E}}.
\newline\urlprefix\url{http://dx.doi.org/10.1039/C6NR05087E}

\bibitem{chen2012}
X.~Chen, C.~K. Wong, C.~A. Yuan, G.~Zhang,
  \href{http://www.sciencedirect.com/science/article/pii/S0925400511010598}{Impact
  of the functional group on the working range of polyaniline as carbon dioxide
  sensors}, Sensors and Actuators B: Chemical 175~(Supplement C) (2012) 15 --
  21, selected Papers presented at Eurosensors XXV.
\newblock \href {http://dx.doi.org/https://doi.org/10.1016/j.snb.2011.11.054}
  {\path{doi:https://doi.org/10.1016/j.snb.2011.11.054}}.
\newline\urlprefix\url{http://www.sciencedirect.com/science/article/pii/S0925400511010598}

\bibitem{chen2012210}
X.~Chen, C.~A. Yuan, C.~K. Wong, H.~Ye, S.~Y. Leung, G.~Zhang,
  \href{http://www.sciencedirect.com/science/article/pii/S0925400512008568}{Molecular
  modeling of protonic acid doping of emeraldine base polyaniline for chemical
  sensors}, Sensors and Actuators B: Chemical 174~(Supplement C) (2012) 210 --
  216.
\newblock \href {http://dx.doi.org/https://doi.org/10.1016/j.snb.2012.08.042}
  {\path{doi:https://doi.org/10.1016/j.snb.2012.08.042}}.
\newline\urlprefix\url{http://www.sciencedirect.com/science/article/pii/S0925400512008568}

\bibitem{zhang2017}
Y.~Zhang, C.~Tan, Q.~Yang, H.~Ye, X.-P. Chen, Arsenic phosphorus monolayer: A
  promising candidate for h 2 s sensor and no degradation with high sensitivity
  and selectivity, IEEE Electron Device Letters 38~(9) (2017) 1321--1324.

\bibitem{hu2017}
F.-F. Hu, H.-Y. Tang, C.-J. Tan, H.-Y. Ye, X.-P. Chen, G.-Q. Zhang, Nitrogen
  dioxide gas sensor based on monolayer sns: A first-principles study, IEEE
  Electron Device Letters 38~(7) (2017) 983--986.

\bibitem{rPBE1998}
Y.~Zhang, W.~Yang,
  \href{https://link.aps.org/doi/10.1103/PhysRevLett.80.890}{Comment on
  ``generalized gradient approximation made simple''}, Phys. Rev. Lett. 80
  (1998) 890--890.
\newblock \href {http://dx.doi.org/10.1103/PhysRevLett.80.890}
  {\path{doi:10.1103/PhysRevLett.80.890}}.
\newline\urlprefix\url{https://link.aps.org/doi/10.1103/PhysRevLett.80.890}

\bibitem{14landauer}
R.~Landauer, \href{http://dx.doi.org/10.1080/14786437008238472}{Electrical
  resistance of disordered one-dimensional lattices}, Philosophical Magazine
  21~(172) (1970) 863--867.
\newblock \href
  {http://arxiv.org/abs/http://dx.doi.org/10.1080/14786437008238472}
  {\path{arXiv:http://dx.doi.org/10.1080/14786437008238472}}, \href
  {http://dx.doi.org/10.1080/14786437008238472}
  {\path{doi:10.1080/14786437008238472}}.
\newline\urlprefix\url{http://dx.doi.org/10.1080/14786437008238472}

\bibitem{15sen}
A.~Sen, C.-C. Kaun, \href{http://dx.doi.org/10.1021/nn101840a}{Effect of
  electrode orientations on charge transport in alkanedithiol single-molecule
  junctions}, ACS Nano 4~(11) (2010) 6404--6408, pMID: 20936842.
\newblock \href {http://arxiv.org/abs/http://dx.doi.org/10.1021/nn101840a}
  {\path{arXiv:http://dx.doi.org/10.1021/nn101840a}}, \href
  {http://dx.doi.org/10.1021/nn101840a} {\path{doi:10.1021/nn101840a}}.
\newline\urlprefix\url{http://dx.doi.org/10.1021/nn101840a}

\bibitem{16fu}
M.-D. Fu, I.-W.~P. Chen, H.-C. Lu, C.-T. Kuo, W.-H. Tseng, C.-h. Chen,
  \href{http://dx.doi.org/10.1021/jp070690u}{Conductance of
  alkanediisothiocyanates:  effect of headgroup−electrode contacts}, The
  Journal of Physical Chemistry C 111~(30) (2007) 11450--11455.
\newblock \href {http://arxiv.org/abs/http://dx.doi.org/10.1021/jp070690u}
  {\path{arXiv:http://dx.doi.org/10.1021/jp070690u}}, \href
  {http://dx.doi.org/10.1021/jp070690u} {\path{doi:10.1021/jp070690u}}.
\newline\urlprefix\url{http://dx.doi.org/10.1021/jp070690u}

\bibitem{17li}
C.~Li, I.~Pobelov, T.~Wandlowski, A.~Bagrets, A.~Arnold, F.~Evers,
  \href{http://dx.doi.org/10.1021/ja0762386}{Charge transport in single au |
  alkanedithiol | au junctions:  coordination geometries and conformational
  degrees of freedom}, Journal of the American Chemical Society 130~(1) (2008)
  318--326, pMID: 18076172.
\newblock \href {http://arxiv.org/abs/http://dx.doi.org/10.1021/ja0762386}
  {\path{arXiv:http://dx.doi.org/10.1021/ja0762386}}, \href
  {http://dx.doi.org/10.1021/ja0762386} {\path{doi:10.1021/ja0762386}}.
\newline\urlprefix\url{http://dx.doi.org/10.1021/ja0762386}

\bibitem{18jang}
S.-Y. Jang, P.~Reddy, A.~Majumdar, R.~A. Segalman,
  \href{http://dx.doi.org/10.1021/nl0609495}{Interpretation of stochastic
  events in single molecule conductance measurements}, Nano Letters 6~(10)
  (2006) 2362--2367, pMID: 17034112.
\newblock \href {http://arxiv.org/abs/http://dx.doi.org/10.1021/nl0609495}
  {\path{arXiv:http://dx.doi.org/10.1021/nl0609495}}, \href
  {http://dx.doi.org/10.1021/nl0609495} {\path{doi:10.1021/nl0609495}}.
\newline\urlprefix\url{http://dx.doi.org/10.1021/nl0609495}

\bibitem{19papa}
T.~A. Papadopoulos, I.~M. Grace, C.~J. Lambert,
  \href{https://link.aps.org/doi/10.1103/PhysRevB.74.193306}{Control of
  electron transport through fano resonances in molecular wires}, Phys. Rev. B
  74 (2006) 193306.
\newblock \href {http://dx.doi.org/10.1103/PhysRevB.74.193306}
  {\path{doi:10.1103/PhysRevB.74.193306}}.
\newline\urlprefix\url{https://link.aps.org/doi/10.1103/PhysRevB.74.193306}

\bibitem{20zhou}
J.~Zhou, B.~Xu, \href{http://dx.doi.org/10.1063/1.3615803}{Determining contact
  potential barrier effects on electronic transport in single molecular
  junctions}, Applied Physics Letters 99~(4) (2011) 042104.
\newblock \href {http://arxiv.org/abs/http://dx.doi.org/10.1063/1.3615803}
  {\path{arXiv:http://dx.doi.org/10.1063/1.3615803}}, \href
  {http://dx.doi.org/10.1063/1.3615803} {\path{doi:10.1063/1.3615803}}.
\newline\urlprefix\url{http://dx.doi.org/10.1063/1.3615803}

\bibitem{21zhou}
J.~Zhou, F.~Chen, B.~Xu, \href{http://dx.doi.org/10.1021/ja900989a}{Fabrication
  and electronic characterization of single molecular junction devices: A
  comprehensive approach}, Journal of the American Chemical Society 131~(30)
  (2009) 10439--10446, pMID: 19722620.
\newblock \href {http://arxiv.org/abs/http://dx.doi.org/10.1021/ja900989a}
  {\path{arXiv:http://dx.doi.org/10.1021/ja900989a}}, \href
  {http://dx.doi.org/10.1021/ja900989a} {\path{doi:10.1021/ja900989a}}.
\newline\urlprefix\url{http://dx.doi.org/10.1021/ja900989a}

\end{thebibliography}

\begin{figure*}
	\centering
	\includegraphics[scale=0.9]{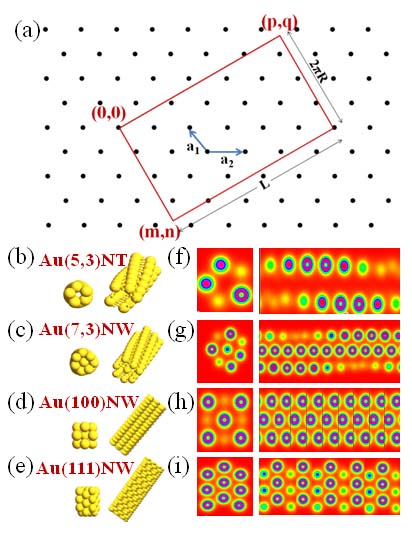}
	\caption{(a) The underlying 2D triangular lattice having basis vectors $a_{1}$ and $a_{2}$, cylindrical folding of which caters to 1D gold (m,n) $chiral$  nanostructures with radius of $(m^{2} + n^{2} - mn)^{1/2}|a_{1}|/2\pi$. Top (left side) and lateral view (right side) for freestanding $chiral$ structures of (b) Au(5,3)NT, (c) Au(7,3)NW, (d) Au(100)NW and (e) Au(111)NW.  The contour plots for electron density of (f) Au(5,3)NT, (g) Au(7,3)NW, (h) Au(100)NW and (i) Au(111)NW  with top as well as lateral view in the same order. The charge density appears to be minimal at the center of Au(5,3)NT, while it is strong in case of Au(7,3)NW, due to presence of the central strand in the latter.}
	\label{fgr:figure1col}
\end{figure*}

\begin{figure*}
	\centering
	\includegraphics[scale=0.5]{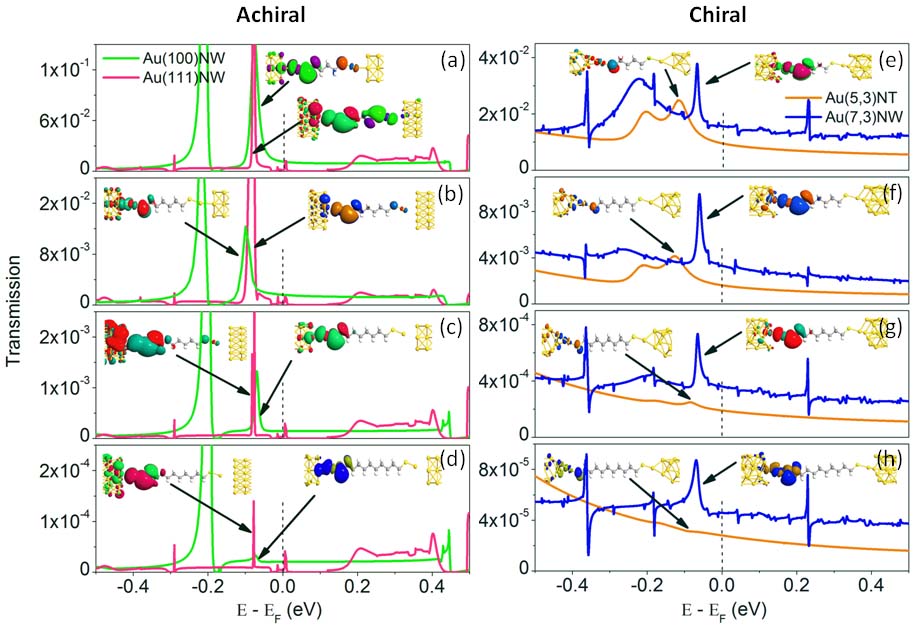}
	\caption{The transmission profile for various alkanedithiol $[ADT, HS-(CH_{2})_{N}-SH]$  single molecule junctions, where the respective molecular moiety bridges either with (a-d) $achiral$ leads of Au(100)NWs and Au(111)NWs or with (e-h) $chiral$ leads of Au(5,3)NTs and Au(7,3)NWs. The molecular chain under study comprises (a,e) butanedithiol, (b,f) hexanedithiol, (c,g) octanedithiol and (d,h) decanedithiol. The arrows indicate the first interacting states of the similar orbital character from the leads with the renormalized molecular levels (RMLs) of the junction molecule. The transmission eigenstates associated with the resonant peak near the Fermi level are shown for comparison with respect to different lead topology. }
	\label{fgr:figure2col}
\end{figure*}

\begin{figure*}
	\centering
	\includegraphics[scale=0.7]{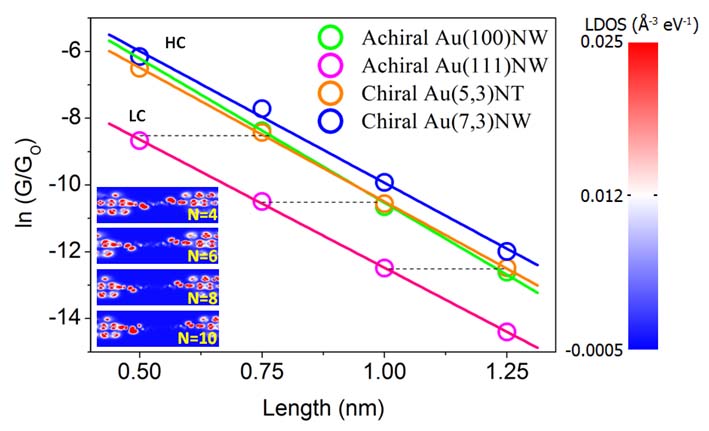}
	\caption{Logarithm of electronic conductance versus molecular length for the $chiral$ Au(5,3)NT (orange) and Au(7,3)NW (blue) based single-molecule junctions (SMJs)  compared to that for the $achiral$ Au(100)NW (green) and Au(111)NW (magenta) based ones. The solid lines refer to the exponential decay in the electronic conductance of SMJs as a function of the molecular length. The horizontal dashed lines demonstrate the similarities in the conductance values for SMJs with different lead structures of varying length. In the inset are shown the contour plots of self-consistent local density of states (LDOS) for $chiral$ Au(7,3)NW based SMJs with the increase in the number (N) of the methylene groups, $CH_{2}$, in the  alkanedithiol chain. 
	}
	\label{fgr:figure3col}
\end{figure*}

\begin{figure*}
	\centering
	\includegraphics[scale=0.9]{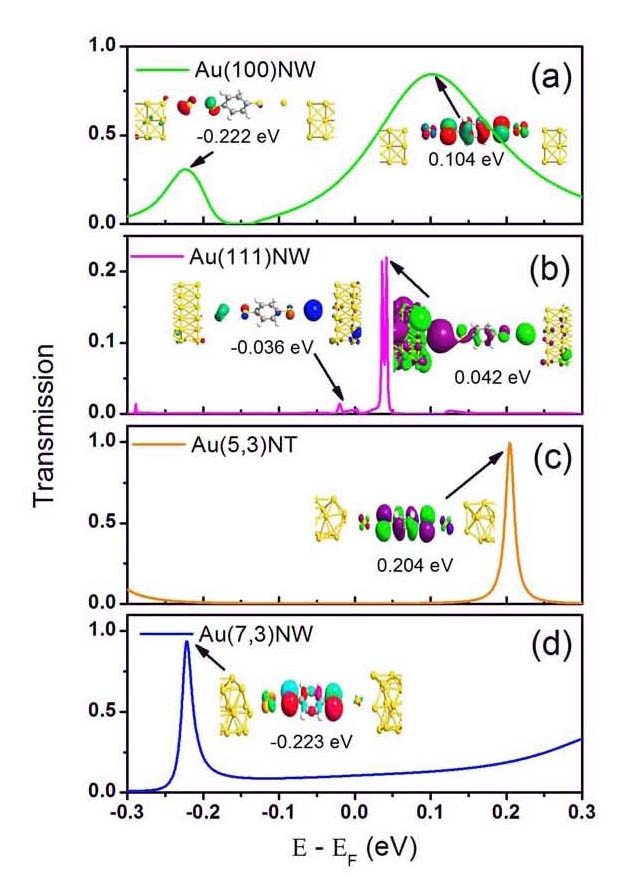}
	\caption{A comparative study of the electronic transmission behavior for various $\pi$-conjugated benzenedithiol single-molecule junctions formed by a set of (a-b) $achiral$ and (c-d) $chiral$ nanoleads. The transmission eigenstates having the isovalue of 0.4, associated with the principal resonant peaks near the Fermi level are also shown.}
	\label{fgr:figure4col}
\end{figure*}

\end{document}